\documentclass[10pt,letterpaper]{article}
\usepackage[T1]{fontenc}
\usepackage[utf8]{inputenc}
\usepackage[margin=1in]{geometry}
\usepackage{array,booktabs,tabularx}
\usepackage{caption}
\usepackage{titlesec}
\titleformat{\section}{\normalfont\large\bfseries}{\thesection}{1em}{}
\usepackage{xurl}
\usepackage[hidelinks,pdfencoding=auto]{hyperref}
\hypersetup{pdftitle={Scarce Inputs and the Growth of Innovative Firms},pdfauthor={Zhu Kei Choo},pdfkeywords={Resource scarcity; innovation; organizational capability; talent allocation; digital matching; infrastructure flexibility}}
\renewcommand{\baselinestretch}{1.25}
\begin{document}
\begin{center}
\vspace*{55pt}
{\LARGE Scarce Inputs and the Growth of Innovative Firms\par}
\vspace{15pt}
{\large Zhu Kei Choo\textsuperscript{*}\par}
\vspace{14pt}
{\normalsize November 2026\par}
\vspace{25pt}
{\small\bfseries Abstract\par}
\end{center}
\vspace{14pt}
\begin{quote}
\small\setlength{\baselineskip}{20.6pt}\noindent
A firm can recruit specialists, buy technology, or secure infrastructure without being able to put those  resources to productive use. An unavailable complement, a poor match, or a decision to wait can explain  the same apparent gap between input and output. For innovation and growth, the crucial question is what  these investments allow a firm to produce, not simply whether it can make them. Drawing on research  on organizational capabilities, skilled labor, digital hiring, and resource-constrained infrastructure, I  compare the decisions behind these outcomes and the alternatives available to firms and workers. Holding  a resource can preserve a valuable future opportunity, but it can also reflect managerial barriers or a  desire to limit a competitor. I discuss how these possibilities might be distinguished in studies of research  projects, external expertise, and infrastructure investment.   

\vspace{8pt}

\end{quote}
\begingroup
\renewcommand{\thefootnote}{\fnsymbol{footnote}}
\footnotetext[1]{Guanghua School of Management, Peking University.}
\endgroup
\clearpage
\section{Introduction}

Innovative firms rarely fail because they lack ideas alone. They need people who can develop those ideas, the knowledge to judge them, the equipment to test them, and an organization willing to commit resources to uncertain projects. When one of these inputs becomes scarce, the most visible response is often to acquire more of it. Yet a firm can hire more researchers without launching more projects, adopt a digital technology without improving its products, or retain access to a data-center site without adding usable computing capacity. These are not necessarily examples of failed investment. They are reasons to examine the steps between obtaining a resource and making productive use of it.

The outcomes are not interchangeable. More researchers or faster recruitment may help a team pursue an invention, but neither tells us whether the invention is completed. A new technology may improve an existing task without yielding a new product, while additional factory or data-center capacity may enlarge output only after it is built and used. I therefore treat employment, operating activity, innovation, and growth as separate outcomes, and ask what connects them in each setting.

A shortage can also be a problem of access or organization, even when the resource exists. A specialist may be hard to find, assigned to work that does not use her expertise, or unable to begin a project because the laboratory budget is missing. A data center, by contrast, may confront an actual limit on locally available water. Increasing supply, improving matching, revising managerial incentives, and investing in complementary equipment address different obstacles. Grouping them all under ``scarcity'' can hide the very margin on which a firm could act.

The research that follows traces this distinction from learning and managerial allocation to recruitment, technology adoption, and physical infrastructure. I am particularly interested in cases where a firm secures a scarce resource but the project that would use it does not move forward. Sometimes that is prudent preparation; sometimes a different allocation might have produced more. Understanding why is the starting point for judging the implications for innovation and growth.

\section{Complementary resources, learning, and organizational choice}

Teece (1986) begins with a distributional puzzle: an inventor does not necessarily capture the returns to an invention. Manufacturing, marketing, distribution, and other complementary assets can determine who benefits from commercialization. This explains why a technically successful innovation may not reward its inventor, but it leaves open how a firm becomes capable of responding to a new opportunity in the first place. Teece, Pisano, and Shuen (1997) shift attention to that problem through dynamic capabilities: firms must integrate and reconfigure competencies as circumstances change. Owning a distribution network may help commercialize one invention without giving the firm the ability to organize around the next.

The difference between possessing an asset and being able to reconfigure it leads naturally to the learning problem developed by Cohen and Levinthal. Their 1989 study argues that internal R\&D not only produces knowledge but also helps firms learn from discoveries made elsewhere. Their 1990 treatment of absorptive capacity goes further, explaining why recognizing, assimilating, and applying external knowledge depend on what an organization already knows. This gives Teece's complementary-assets argument an important qualification. Buying a technology or recruiting its inventor may secure access, but a firm that has abandoned its own exploratory work could find itself less able to assess and use what it has acquired. The missing complement may be knowledge within the firm rather than a physical asset outside it.

Knowledge within the firm is still not enough if managers do not allocate people and budgets to make use of it. Sirmon, Hitt, and Ireland (2007) distinguish assembling resources from building capabilities and putting them into use; Sirmon et al. (2011) examine how these choices vary across organizational levels and stages of development. Their work moves the discussion from what a firm knows to what it chooses to do. A company may have the expertise to undertake a project but leave that expertise underused because funding, authority, or coordination is missing. That observation also suggests why an apparently scarce skill may be plentiful on paper yet unavailable to the team that needs it.

Not all of those complements must be supplied internally. Forman, Goldfarb, and Greenstein (2008) examine advanced Internet investment and find that local resources and internal firm capabilities have to be considered together. Access to outside expertise in a city can substitute for some internal inputs, whereas a firm operating where those services are scarce may depend more heavily on its own resources. The finding extends the absorptive-capacity argument without overturning it: outside help can ease a constraint, but whether a firm can benefit from that help still depends on the task and its capabilities. Baker and Nelson (2005) describe another way around a missing input, showing how entrepreneurial bricolage recombines resources that others overlook. Such improvisation can change how much of a resource a project requires; it cannot, however, make every physical limit disappear.

There is also a reason for underuse that does not involve a missing complement at all. March (1991) shows how the balance between exploration and exploitation can favor established activities over uncertain new ones. This helps connect the capability literature to the choices discussed next: even a firm that could deploy a specialist may choose not to fund the specialist's proposed project. The distinction matters because a technical inability to proceed, a managerial reluctance to proceed, and a deliberate decision to wait imply different remedies.

\section{Specialized labor: productive retention or displaced innovation?}

The distinction between capability and choice becomes particularly difficult to observe with research workers. Specialized knowledge takes years to acquire, and employment records say little about whether an employee's best ideas are being pursued. Bidwell (2011), comparing external hires and internal movers in investment banking, documents differences in pay, performance, and exits. Although this is not a study of laboratory scientists, it shows why a long-standing internal match may carry valuable firm-specific knowledge. Retaining a worker is therefore not by itself evidence that management is blocking innovation; replacing that worker could destroy a productive match.

The same advantage of an internal match can create a conflict over who gets to use the worker. Haegele (2026) finds that managers' talent hoarding deters employees from applying for positions elsewhere within the firm, limiting career progression and changing the allocation of workers across teams. A manager has reason to keep a strong employee, even when another department might benefit more from that person's skills. Akcigit and Goldschlag (2023) take the allocation question across firm boundaries by tracing US inventors employed by established and younger firms. Their analysis raises the possibility that higher pay at incumbents does not coincide with higher measured inventive output. But the comparison also introduces a harder counterfactual than the internal-mobility case: researchers choose employers, and alternative firms may lack the resources needed to carry out the same work. Differences between employer groups cannot reveal what that same inventor would have produced elsewhere---the comparison needed to claim a loss of innovation.

Wu and Qian (2025) sharpen the interfirm concern by examining why an incumbent might benefit from keeping researchers even when it is not expanding complementary research activity at the same pace. They report a 0.31-percentage-point increase in affected firms' monthly net hiring following a change associated with patent eligibility, driven largely by fewer separations. Read alongside Akcigit and Goldschlag, this provides a reason to look at retention rather than recruitment alone. It does not establish that the retained researchers are idle or that competitors would have innovated more: the same legal change may affect the profitability of the incumbent's own projects. Haegele's managerial conflict and Wu and Qian's interfirm incentives consequently require different counterfactual comparisons, even though both can leave a skilled worker in the same job.

There is a further reason not to equate employment at an incumbent with lost entrepreneurship. Babina and Howell (2024) study how corporate R\&D contributes to ventures founded by employees, including ventures outside the parent's industry. If incumbent research helps workers acquire ideas and skills that they later carry into new firms, a period of stable employment can support rather than suppress entry. Together, these studies make retention a question about the location and timing of inventive work, not a proxy for innovative success. Knowledge developed at an incumbent may later support a new venture, while a blocked move may leave a better project untried.

\section{When does retaining an option become inefficient?}

Strategic reserve capacity creates a difficult boundary for claims about hoarding. A researcher may be waiting for complementary equipment, regulatory approval, or a technical breakthrough that cannot be scheduled in advance. Keeping that researcher employed can preserve the option to act quickly when a project becomes feasible. March (1991) helps explain why exploration and routine production have different time profiles. Whether the eventual project succeeds is not the same question as whether retaining the researcher was a sensible decision when the firm faced uncertainty.

Evidence of an actual research program, rather than headcount alone, makes the preparation explanation more convincing. Yet the outcome of that program cannot settle whether the decision was sensible when it was made: an uncertain project may fail despite prudent preparation. The more revealing question is whether the firm's other feasible uses of the researcher's time were considered.

\section{Digital participation is not a labor-market outcome}

A firm can report difficulty hiring even when suitable workers are available. The obstacle may be that employers and applicants do not find each other, or that applicants cannot present their experience in ways employers recognize. Kanfer, Wanberg, and Kantrowitz (2001) show that job search reflects motivation and self-regulation as well as opportunity. A person in stable employment may search less because there is little reason to leave, not because online platforms have failed.

The gap between being available for work and being visible to an employer appears in evidence on online job search. Dillahunt et al. (2021) document variation in how US job seekers use online services and complete applications. Their findings show where an application process can be difficult, but they do not establish the causal effect of platform use on employment. Wu, Dong, and He (2026) study a related, broader question by associating digital engagement with labor-market circumstances. More frequent internet use may accompany better job opportunities, yet it can also reflect existing skills, income, or employment status. Read together, these studies caution against treating either platform access or internet-use frequency as a measure of the workers a firm can actually recruit.

Experiments with recruitment tools help locate the friction more precisely. Horton (2017) tests algorithmic recommendations for technical vacancies and finds that treated employers are more likely to fill them. Wiles, Munyikwa, and Horton (2025) instead examine assistance with how applicants present their experience in resumes. Where Horton changes which candidates employers encounter, Wiles and coauthors address whether employers can recognize a candidate's qualifications from the application. Both interventions can improve matching without increasing the underlying supply of skilled people. A faster hire for one firm may even mean a longer search for another.

Better digital access also need not distribute gains evenly. Forman, Goldfarb, and Greenstein (2012), extending their earlier work on internal and local resources, find that investment in advanced Internet technology is associated with substantial wage growth primarily in US counties that were already wealthy, educated, and intensive users of information technology. This is a regional wage result, not a direct estimate of an individual firm's innovation. It does show why removing a digital access barrier may yield much larger economic returns where skilled workers and complementary capabilities are already concentrated. The question of which candidates become visible is not confined to vacancy recommendations or resumes. Rajkumar et al. (2022) use randomized changes to LinkedIn recommendations to show that network ties affect job transmission, making the set of opportunities a worker encounters partly a product of platform design. Better exposure is not automatically better screening, however. Raghavan et al. (2020) show why claims about algorithmic hiring have to be evaluated in light of the target being predicted, the validation procedure, and the errors a system makes. These studies separate a shortage of qualified workers from a failure to identify them. Better matching matters for innovation when it brings together people and projects that otherwise would have been delayed or abandoned. Filling a vacancy more quickly is evidence about the match, not yet about the invention; the next question is whether the new employee joins productive work that the firm can actually support.

\section{Technology adoption, complementary investment, and firm growth}

Finding a suitable employee only solves part of the problem. The firm still has to organize work so that the employee, and the technologies used alongside them, can contribute to production. Brynjolfsson and Hitt (2000) emphasize that information technology often requires changes in work practices and organizational structure. Brynjolfsson, Rock, and Syverson (2021) further explain why investments in intangible complements can make early productivity measurements misleading. A firm may incur training and reorganization costs before gains appear, although delayed gains cannot be assumed whenever performance is initially weak.

The long-standing argument about organizational complements becomes more concrete when a new tool changes a worker's daily tasks. Brynjolfsson, Li, and Raymond (2025) study the introduction of a generative-AI assistant in customer service and report higher issues resolved per hour, with gains varying by experience and skill. In this setting the tool can help less experienced workers draw on established knowledge. The result is informative precisely because the outcome is close to the task being changed. It does not mean the same tool would help a scientist whose constraint is an experiment that has not been funded, nor does task productivity alone demonstrate product innovation.

A task-level gain becomes an innovation question when a firm uses the saved time or new capability to develop something it could not previously offer. Babina, Fedyk, He, and Hodson (2024) construct a measure of AI investment from employee resumes and examine sales, employment, firm value, and product innovation. Their empirical strategy offers more than a raw comparison of technology adopters and nonadopters, although the interpretation continues to depend on the AI-investment measure and the identifying assumptions. Their finding of concentrated gains among larger firms also qualifies what the customer-service evidence can tell us. Better performance in a particular activity may support new products, but firm growth may also reflect customers moving from competitors. A larger adopting firm is not necessarily evidence of a larger market.

That distinction between growth within an incumbent and growth elsewhere brings the analysis back to corporate research. Babina and Howell (2024) follow employee-founded ventures that benefit from knowledge developed within established companies. Whereas Babina et al. (2024) study how an incumbent's own AI investment relates to its products and performance, Babina and Howell show why the returns to an incumbent's research can also emerge in new firms. The comparison broadens the accounting of innovation: a corporate research program can matter for growth outside the company that finances it.

\section{Infrastructure scarcity and choices already embedded in assets}

The importance of complementary investment is especially visible when a firm cannot readily replace the assets it has already installed. A data-center operator chooses its site, power connections, buildings, and cooling systems before it knows precisely how computing demand or local water availability will change. Fine and Freund (1990) and Van Mieghem (1998) show, in models of flexible capacity, why paying for alternative production possibilities can be worthwhile under uncertainty. Sanchez (1995) adds a useful distinction between possessing flexible resources and coordinating their use. The issue therefore shifts from whether a firm can acquire an input today to whether past investments have left it any workable options when the input becomes scarce.

A constraint may sometimes make new technology worth developing, rather than merely force a firm to curtail production. Popp (2002) documents how higher energy prices help direct inventive activity toward energy-saving technologies. Popp (2006) then examines patent evidence from the United States, Japan, and Germany to show how pollution-control regulation relates to both domestic invention and international technology diffusion. Price incentives and regulation differ from a local interruption of water supply, but these studies add an important possibility: a constraint may change the direction of innovation itself, not only which existing technology a firm deploys. The response still depends on whether firms can develop, acquire, or learn to use the alternative.

Qian and Wu (2026) examine California data centers around the tightening of water conditions in Santa Clara County. Their facility panel covers 90 data centers during 2018--2025, including 41 in Santa Clara. The average electricity response is imprecisely estimated. The authors' triple-difference estimate implies a 192.4\% relative change in the ratio of electricity use at cooling-flexible versus less flexible facilities in Santa Clara, compared with the corresponding ratio elsewhere in California. They also estimate a 15.58-percentage-point differential in the probability of a documented capacity-expansion project in the sample with pre- and post-shock electricity records. They indicate where operators could sustain activity and sought to add capacity under the water constraint.

The evidence stops short of establishing how much new computing service was delivered. Electricity meters record cooling as well as computing, project records can precede completed construction, and installed cooling systems reflect earlier choices rather than random assignment. Those limitations do not erase the observed divergence. They identify the missing step between an operator seeking to expand and a customer receiving additional usable capacity.

A comparison with electricity shortages shows both the appeal and the limits of this argument. Allcott, Collard-Wexler, and O'Connell (2016) study Indian manufacturers that can partly offset grid shortages with generators; generator costs and economies of scale affect which firms can do so. As with cooling flexibility, an installed alternative can reduce vulnerability to an input shock, but the technology and cost structures differ. Kogut and Kulatilaka (1994) examine yet another option---shifting production across a multinational network---and their 2001 study relates such flexibility to organizational capabilities. Moving output between sites is not the same as switching a cooling mode within a fixed site. Fang and Greenstein (2025) add information about why cloud providers and third-party data-center operators locate differently, but these location patterns do not measure the cost of relocation or the feasibility of transferring computing workloads. The papers are most useful in combination when their different margins of adjustment remain visible.

Flexibility can also move a constraint rather than eliminate it. Siddik, Shehabi, and Marston (2021) account for data centers' direct water use as well as water consumed in electricity generation. This complicates the interpretation of an operator's response to scarcity: a cooling choice that protects on-site activity may increase electricity demand or shift water use elsewhere. Qian and Wu's facility evidence concerns how sites adjust; Siddik and coauthors remind us that a privately useful adjustment need not reduce the total environmental burden. Thus an investment that helps a site expand can improve the operator's growth prospects without necessarily reducing the resource costs of that growth.

\section{When holding resources makes sense---and when it does not}

Much of the difficulty in comparing these papers comes from the outcomes they observe. Haegele (2026) measures barriers to moving within a firm; Horton (2017) measures whether vacancies are filled; Babina et al. (2024) get closer to the products and performance promised in this paper's title. Table 1 keeps those differences visible rather than treating each estimate as another measure of growth.

\clearpage
\begin{table}[!htbp]
\centering
\caption{What the reviewed studies identify---and what remains unobserved}
\label{tab:studies}
\setlength{\tabcolsep}{3pt}
\renewcommand{\baselinestretch}{1}
\fontsize{7}{10.9}\selectfont
\renewcommand{\arraystretch}{1.05}
\begin{tabularx}{\textwidth}{@{}>{\raggedright\arraybackslash}p{78pt}>{\raggedright\arraybackslash}p{114pt}>{\raggedright\arraybackslash}p{121pt}>{\raggedright\arraybackslash}X@{}}
\toprule
\textbf{Study} & \textbf{Method and main outcome} & \textbf{Main identification or interpretation limit} & \textbf{Unmeasured downstream outcome} \\
\midrule
Cohen \& Levinthal (1989, 1990) & Conceptual analysis of R\&D, learning and absorptive capacity & These papers develop a mechanism; they do not identify a common causal effect for the other settings & Project-level commercialization from acquiring expertise \\
Akcigit \& Goldschlag (2023) & Employment histories and inventor outcomes across employers & Sorting and unobserved alternative projects complicate comparisons & Counterfactual innovation by the same inventor \\
Haegele (2026) & Personnel and quasi-random evidence that hoarding deters internal job applications & Mobility effects need not equal firm-wide output effects & Projects completed after a different internal assignment \\
Babina et al. (2024) & Firm-level evidence on AI investment, product innovation and growth & Firm-level gains need not be net market gains & Displacement-adjusted market-wide new products \\
Horton (2017) & Field experiment on recommendations and technical-vacancy filling & Treatment is evaluated for a platform and selected vacancies & Net employment creation and long-run match quality \\
Brynjolfsson et al. (2025) & Workplace rollout and support-agent productivity & Task-specific effects do not directly identify product innovation & New products and subsequent firm growth \\
Qian \& Wu (2026) & Facility comparisons of cooling, electricity use and project records & Installed equipment and IT demand are not randomized or directly observed & Commissioned IT capacity and economic output \\
\bottomrule
\end{tabularx}
\par\smallskip
\textit{Notes:} The studies differ in methods and outcome measures. The final column identifies outcomes not directly observed in the cited study.
\end{table}

The common thread across these findings is not a single form of scarcity. In the labor studies, the relevant choice is who can work on a project: a manager may prevent an internal transfer, while a company may retain an employee who could join a rival. In the infrastructure study, an operator has already committed to a site and must decide how far to expand it. All three decisions can change which projects proceed, but the consequences occur at different levels. Blocking an internal transfer limits the opportunities of another team in the same company; retaining a researcher against outside offers may affect a competing firm; choosing an installed cooling system changes what an existing facility can operate or expand. None of these outcomes is, on its own, a measure of lost innovation.

The distinction also changes what we would expect to observe. When outside demand for a specialized researcher increases, an employer with functioning laboratories, funding, and suitable projects can respond by both retaining and deploying the worker. An employer lacking those complements may increase retention without expanding research activity. That comparison would be most informative among firms facing similar shifts in demand for their researchers but differing in their ability to finance and organize new work.

A stronger claim, that retaining the researcher reduces innovation elsewhere, requires evidence about the alternative employer. A competitor might have a viable project, but it might also lack complementary resources or pursue a discovery that the incumbent would have made anyway. It is possible for retaining talent to be privately sensible and for an alternative allocation to generate more innovation; it is not enough to infer that possibility from a high salary or a low contemporaneous patent count. This is the point at which the labor literature poses a question that the cooling-flexibility case does not: not only whether a firm grows, but whether its control of a scarce input changes the growth opportunities of its rivals.

\section{Alternative explanations and the scope of the argument}

The central uncertainty is the value of work that never occurred. An employee who stays at an incumbent might have led a successful project at a young firm, but an incumbent's equipment, collaborators, and accumulated knowledge may also have made the employee's eventual contribution possible. Inventor employment and patents help identify patterns; they cannot alone recover either unrealized outcome. This is why the allocation question needs a counterfactual, rather than a judgment inferred from a period of low observed output.

Organizational scale complicates the comparison further. A large firm may have more projects in which a scientist can eventually contribute, but it may also be better able to retain that scientist when a smaller rival needs the same expertise. Akcigit and Goldschlag (2023) study inventor allocation and innovative output, while Babina et al. (2024) find that the benefits associated with AI investment are concentrated among larger firms. The papers do not identify the same mechanism, but together they show why a scale advantage should not automatically be read as either more efficient deployment or exclusionary conduct. Direct evidence about research assignments and products is needed to tell the stories apart.

A final distinction concerns whose growth is being measured. A successful hire may move a worker from one employer to another, just as new sales at an AI-adopting firm may come partly from competitors. Neither transfer should be confused with no benefit: improved matching or competition can still raise productivity. But claims about aggregate innovation require evidence that new products, research output, or usable capacity were created rather than only moved between firms.

Finally, flexibility need not improve welfare simply because it protects a firm's operations. Switching cooling modes can shift water demand from a facility to the generation of the electricity it buys, an issue raised by Siddik et al. (2021). Moving production across countries, as in Kogut and Kulatilaka (1994), entails different constraints and external effects. The lesson is not that resource flexibility is bad; it is that private growth and the gains to other firms, workers, or communities should be evaluated separately.

\section{Questions worth pursuing}

The next step is to follow resource decisions into outcomes that can establish whether innovation took place and whether the resulting growth was additional. Three comparisons seem especially useful.

The question I find most promising is whether competition for a particular scientific skill changes what researchers are assigned to do. Patent histories and inventor employment records can reveal where researchers move and what they patent afterward. Combined with firm R\&D spending and dated job advertisements, those sources could show whether employers exposed to stronger outside competition retain more people and whether research output changes later. They cannot, however, tell us whether an employee spent the intervening years on an unfunded proposal, a difficult but valuable experiment, or routine work. That distinction would require project assignments and budgets from cooperating firms.

A useful comparison would follow employers that faced similar increases in outside demand for a skill but differed beforehand in their ability to support research using that skill. If firms with fewer complementary capabilities retained more researchers without adding funded projects, the result would fit the account developed here. If those same firms increased research deployment just as strongly, it would weaken it. The design would also need to distinguish internal moves from departures, and to check whether the outside-demand change coincided with better commercial prospects for the research itself. Most importantly, even a gap between retention and projects would not by itself show that a rival lost an innovation: that requires evidence of a credible alternative use.

A second question concerns firms that try to overcome a missing internal capability by buying expertise from outside. Forman, Goldfarb, and Greenstein (2008) suggest why the answer may depend on both location and what the firm already knows. One could compare subsequent products or processes among firms with different prior technical experience as specialized services become more accessible, using patent records and company accounts before seeking contract-level evidence. The challenge is that capable firms may have located near these services long before access improved. The interesting result would therefore be a change in innovation following new access, not simply a difference between firms in cities with more specialists.

For infrastructure, I would follow expansion records through commissioning and actual computing service rather than stop at electricity consumption or permits. Public energy reports and project filings could be paired with verified commissioning dates, while pre-shock differences in customers, building age, and power access would remain important competing explanations. If flexible sites ultimately deliver more computing service, that strengthens the connection to firm growth; comparing their gains with changes at the operator's other sites would then help separate new activity from relocation.

These studies will also require patience about timing. Brynjolfsson et al. (2021) explain why productivity gains can lag the organizational investments needed to achieve them. But longer follow-up is useful only if it records the eventual project, product, or operating capacity and, where possible, the activity displaced elsewhere.

\section{Conclusion}

My strongest conclusion concerns research talent. When two firms compete for a specialist, the decisive issue is not which one can pay more or keep the worker longer, but which has the capabilities and viable projects to make good use of that expertise. Retention may be prudent under uncertainty, yet it warrants closer scrutiny when a feasible project elsewhere is foreclosed and no comparable work is being supported internally. We cannot settle that question from headcounts or patents alone.

The other settings show that different constraints call for different responses. Digital matching can bring a worker to a project, but local capabilities shape the returns to digital adoption. Infrastructure flexibility can keep a site productive during a resource shock, while the price and regulatory pressures examined by Popp (2002, 2006) can also encourage firms to invent or adopt alternatives. The routes to growth differ, and so do the outcomes that would establish it. A convincing account of scarce inputs should follow these choices to completed innovations and additional output while asking which opportunities were displaced. That is a harder question than identifying the firm that secured the resource, but it is the one most worth answering.

\clearpage

\section*{References}

\addcontentsline{toc}{section}{References}

\begingroup

\renewcommand{\baselinestretch}{1}
\fontsize{8}{15.3}\selectfont

\setlength{\parindent}{0pt}

\setlength{\parskip}{6pt}

\hangindent=8pt\hangafter=1\noindent Akcigit, U., \& Goldschlag, N. (2023). Where have all the ``creative talents'' gone? Employment dynamics of US inventors. NBER Working Paper No. 31085. \url{https://doi.org/10.3386/w31085}\par

\hangindent=8pt\hangafter=1\noindent Allcott, H., Collard-Wexler, A., \& O'Connell, S. D. (2016). How do electricity shortages affect industry? Evidence from India. American Economic Review, 106(3), 587--624. \url{https://doi.org/10.1257/aer.20140389}\par

\hangindent=8pt\hangafter=1\noindent Babina, T., Fedyk, A., He, A., \& Hodson, J. (2024). Artificial intelligence, firm growth, and product innovation. Journal of Financial Economics, 151, Article 103745. \url{https://doi.org/10.1016/j.jfineco.2023.103745}\par

\hangindent=8pt\hangafter=1\noindent Babina, T., \& Howell, S. T. (2024). Entrepreneurial spillovers from corporate R\&D. Journal of Labor Economics, 42(2), 469--509. \url{https://doi.org/10.1086/723501}\par

\hangindent=8pt\hangafter=1\noindent Baker, T., \& Nelson, R. E. (2005). Creating something from nothing: Resource construction through entrepreneurial bricolage. Administrative Science Quarterly, 50(3), 329--366. \url{https://doi.org/10.2189/asqu.2005.50.3.329}\par

\hangindent=8pt\hangafter=1\noindent Bidwell, M. (2011). Paying more to get less: The effects of external hiring versus internal mobility. Administrative Science Quarterly, 56(3), 369--407. \url{https://doi.org/10.1177/0001839211433562}\par

\hangindent=8pt\hangafter=1\noindent Brynjolfsson, E., \& Hitt, L. M. (2000). Beyond computation: Information technology, organizational transformation and business performance. Journal of Economic Perspectives, 14(4), 23--48. \url{https://doi.org/10.1257/jep.14.4.23}\par

\hangindent=8pt\hangafter=1\noindent Brynjolfsson, E., Li, D., \& Raymond, L. (2025). Generative AI at work. The Quarterly Journal of Economics, 140(2), 889--942. \url{https://doi.org/10.1093/qje/qjae044}\par

\hangindent=8pt\hangafter=1\noindent Brynjolfsson, E., Rock, D., \& Syverson, C. (2021). The productivity J-curve: How intangibles complement general purpose technologies. American Economic Journal: Macroeconomics, 13(1), 333--372. \url{https://doi.org/10.1257/mac.20180386}\par

\hangindent=8pt\hangafter=1\noindent Cohen, W. M., \& Levinthal, D. A. (1989). Innovation and learning: The two faces of R \& D. The Economic Journal, 99(397), 569--596. \url{https://doi.org/10.2307/2233763}\par

\hangindent=8pt\hangafter=1\noindent Cohen, W. M., \& Levinthal, D. A. (1990). Absorptive capacity: A new perspective on learning and innovation. Administrative Science Quarterly, 35(1), 128--152. \url{https://doi.org/10.2307/2393553}\par

\hangindent=8pt\hangafter=1\noindent Dillahunt, T. R., Israni, A., Lu, A. J., Cai, M., \& Hsiao, J. C.-Y. (2021). Examining the use of online platforms for employment: A survey of U.S. job seekers. In Proceedings of the 2021 CHI Conference on Human Factors in Computing Systems (Article 562, pp. 1--23). ACM. \url{https://doi.org/10.1145/3411764.3445350}\par

\hangindent=8pt\hangafter=1\noindent Fang, T. P., \& Greenstein, S. (2025). Where the cloud rests: The economic geography of data centers. Strategy Science, 10(4), 404--420. \url{https://doi.org/10.1287/stsc.2024.0225}\par

\hangindent=8pt\hangafter=1\noindent Fine, C. H., \& Freund, R. M. (1990). Optimal investment in product-flexible manufacturing capacity. Management Science, 36(4), 449--466. \url{https://doi.org/10.1287/mnsc.36.4.449}\par

\hangindent=8pt\hangafter=1\noindent Forman, C., Goldfarb, A., \& Greenstein, S. (2008). Understanding the inputs into innovation: Do cities substitute for internal firm resources? Journal of Economics \& Management Strategy, 17(2), 295--316. \url{https://doi.org/10.1111/j.1530-9134.2008.00179.x}\par

\hangindent=8pt\hangafter=1\noindent Forman, C., Goldfarb, A., \& Greenstein, S. (2012). The internet and local wages: A puzzle. American Economic Review, 102(1), 556--575. \url{https://doi.org/10.1257/aer.102.1.556}\par

\hangindent=8pt\hangafter=1\noindent Haegele, I. (2026). Talent hoarding in organizations. American Economic Review, 116(8), 3110--3151. \url{https://doi.org/10.1257/aer.20220264}\par

\hangindent=8pt\hangafter=1\noindent Horton, J. J. (2017). The effects of algorithmic labor market recommendations: Evidence from a field experiment. Journal of Labor Economics, 35(2), 345--385. \url{https://doi.org/10.1086/689213}\par

\hangindent=8pt\hangafter=1\noindent Kanfer, R., Wanberg, C. R., \& Kantrowitz, T. M. (2001). Job search and employment: A personality-motivational analysis and meta-analytic review. Journal of Applied Psychology, 86(5), 837--855. \url{https://doi.org/10.1037/0021-9010.86.5.837}\par

\hangindent=8pt\hangafter=1\noindent Kogut, B., \& Kulatilaka, N. (1994). Operating flexibility, global manufacturing, and the option value of a multinational network. Management Science, 40(1), 123--139. \url{https://doi.org/10.1287/mnsc.40.1.123}\par

\hangindent=8pt\hangafter=1\noindent Kogut, B., \& Kulatilaka, N. (2001). Capabilities as real options. Organization Science, 12(6), 744--758. \url{https://doi.org/10.1287/orsc.12.6.744.10082}\par

\hangindent=8pt\hangafter=1\noindent March, J. G. (1991). Exploration and exploitation in organizational learning. Organization Science, 2(1), 71--87. \url{https://doi.org/10.1287/orsc.2.1.71}\par

\hangindent=8pt\hangafter=1\noindent Popp, D. (2002). Induced innovation and energy prices. American Economic Review, 92(1), 160--180. \url{https://doi.org/10.1257/000282802760015658}\par

\hangindent=8pt\hangafter=1\noindent Popp, D. (2006). International innovation and diffusion of air pollution control technologies: The effects of NOX and SO2 regulation in the US, Japan, and Germany. Journal of Environmental Economics and Management, 51(1), 46--71. \url{https://doi.org/10.1016/j.jeem.2005.04.006}\par

\hangindent=8pt\hangafter=1\noindent Qian, Z., \& Wu, S. (2026). How data centers grow when water is scarce. SSRN. \url{https://doi.org/10.2139/ssrn.7599482}\par

\hangindent=8pt\hangafter=1\noindent Raghavan, M., Barocas, S., Kleinberg, J., \& Levy, K. (2020). Mitigating bias in algorithmic hiring: Evaluating claims and practices. In Proceedings of the 2020 Conference on Fairness, Accountability, and Transparency (pp. 469--481). ACM. \url{https://doi.org/10.1145/3351095.3372828}\par

\hangindent=8pt\hangafter=1\noindent Rajkumar, K., Saint-Jacques, G., Bojinov, I., Brynjolfsson, E., \& Aral, S. (2022). A causal test of the strength of weak ties. Science, 377(6612), 1304--1310. \url{https://doi.org/10.1126/science.abl4476}\par

\hangindent=8pt\hangafter=1\noindent Sanchez, R. (1995). Strategic flexibility in product competition. Strategic Management Journal, 16(S1), 135--159. \url{https://doi.org/10.1002/smj.4250160921}\par

\hangindent=8pt\hangafter=1\noindent Siddik, M. A. B., Shehabi, A., \& Marston, L. (2021). The environmental footprint of data centers in the United States. Environmental Research Letters, 16(6), Article 064017. \url{https://doi.org/10.1088/1748-9326/abfba1}\par

\hangindent=8pt\hangafter=1\noindent Sirmon, D. G., Hitt, M. A., \& Ireland, R. D. (2007). Managing firm resources in dynamic environments to create value: Looking inside the black box. Academy of Management Review, 32(1), 273--292. \url{https://doi.org/10.5465/amr.2007.23466005}\par

\hangindent=8pt\hangafter=1\noindent Sirmon, D. G., Hitt, M. A., Ireland, R. D., \& Gilbert, B. A. (2011). Resource orchestration to create competitive advantage: Breadth, depth, and life cycle effects. Journal of Management, 37(5), 1390--1412. \url{https://doi.org/10.1177/0149206310385695}\par

\hangindent=8pt\hangafter=1\noindent Teece, D. J. (1986). Profiting from technological innovation: Implications for integration, collaboration, licensing and public policy. Research Policy, 15(6), 285--305. \url{https://doi.org/10.1016/0048-7333(86)90027-2}\par

\hangindent=8pt\hangafter=1\noindent Teece, D. J., Pisano, G., \& Shuen, A. (1997). Dynamic capabilities and strategic management. Strategic Management Journal, 18(7), 509--533. \url{https://doi.org/10.1002/%28SICI%291097-0266%28199708%2918%3A7%3C509%3A%3AAID-SMJ882%3E3.0.CO%3B2-Z}\par

\hangindent=8pt\hangafter=1\noindent Van Mieghem, J. A. (1998). Investment strategies for flexible resources. Management Science, 44(8), 1071--1078. \url{https://doi.org/10.1287/mnsc.44.8.1071}\par

\hangindent=8pt\hangafter=1\noindent Wiles, E., Munyikwa, Z., \& Horton, J. J. (2025). Algorithmic writing assistance on jobseekers' resumes increases hires. Management Science, 71(12), 10144--10164. \url{https://doi.org/10.1287/mnsc.2024.04528}\par

\hangindent=8pt\hangafter=1\noindent Wu, S., Dong, Y. R., \& He, S. (2026). Digital engagement, income disparities, and job-seeking in the United States since 2010. In Proceedings of the 2026 International Conference on Information Technology for Social Good (GoodIT '26) (9 pp.). ACM. \url{https://doi.org/10.1145/3794786.3830729}\par

\hangindent=8pt\hangafter=1\noindent Wu, S., \& Qian, Z. (2025). Talent hoarding and upstream innovation: Labor market distortions by large incumbents. SSRN. \url{https://doi.org/10.2139/ssrn.5127023}\par

\endgroup

\end{document}